\documentclass[12pt]{iopart}

\usepackage{graphicx} 
\usepackage[backend=biber,natbib=true, citestyle=numeric-comp,sorting=none,defernumbers=true,bibencoding=utf8,style=numeric]{biblatex}

\usepackage{color}
\begin{document}

\title[ Route to Achieving Enhanced  Quantum Capacitance ..]{ Route to Achieving Enhanced  Quantum Capacitance in Functionalized Graphene based
	Supercapacitor Electrodes }
\author{Sruthi T \& Kartick Tarafder}
\address{Department of Physics, National Institute of Technology, Srinivasnagar, Surathkal, Mangalore Karnataka-575025, India}
\ead{karticktarafder@gmail.com}
\vspace{10pt}
\begin{indented}
\item[]\today
\end{indented}
\begin{abstract}
We have investigated the quantum capacitance ($C_Q$) in  functionalized graphene, modified with ad-atoms from different groups in the periodic table.
  Changes in the electronic band structure of graphene upon functionalization and subsequently the quantum capacitance ($C_Q$) of the modified graphene were  systematically analyzed using  density functional theory(DFT) calculations. We observed that the quantum capacitance can be enhanced significantly by means of controlled doping of N, Cl and P ad-atoms in the pristine graphene surface. These ad-atoms are behaving as magnetic impurities in the system, generates a localized density of states near the Fermi energy, which intern  increases charge(electron/hole) carrier density in the system. As a result, a very high quantum capacitance was observed. Finally, the temperature dependent study of  $C_Q$ for  Cl and N functionalized graphene  shows that the $C_Q$ remains very high in a wide range of temperature  near the room temperature. 
\end{abstract}

\section{Introduction}

Large scale generation of green energy from renewable energy sources is utmost necessary in the current scenario. Sunlight is the most viable renewable energy source present in our planet. However energy cannot be produced from the sun uniformly all the time through out the year in many parts of the glob. Therefore, an efficient storage for generated energy and its cost effective transportation is very essential. Hence the designing of efficient energy storage devices is one of the active area of research in green energy production. Supercapacitors based on two-dimensional materials would be a promising technology may provide conceivable alternative for the energy storage \cite{1}.
As basic requirements, supercapacitor should have a very large ion-density, fast charging and discharging capacity with long life time. Two dimensional (2D) materials could play an important role to design an efficient supercapacitor electrodes. With a very high surface area, conductivity and mechanical robustness in 2D materials, specially  functionalized graphenes could be the best choice for supercapacitor electrodes \cite{2,3,4}. The total capacitance ($C_{T}$) of a supercapacitor depends on two component \cite{5}, namely the electrical double layer capacitance ($C_D$) and the quantum capacitance (${C_Q}$) such that .

\[ \frac{1}{C_{T}}=\frac{1}{C_D} +\frac{1}{C_Q}  \]

Insufficiency in either of them will reduce the total capacitance of the device. Thus electrode materials with sufficiently large quantum capacitance is an obligatory factor to obtain high energy density. The quantum capacitance part of electrodes depend on the electronic structure of the electrode materials\cite{6,7}. In case of pristine graphene, the quantum capacitance is very small\cite{8}. However the capacitance can be enhance in graphene based electrode by introducing vacancy defect as well as doped with impurities in a control manner\cite{9}. Nitrogenation and chlorination of graphene could be an effective way to improve the quantum capacitance in the system \cite{10,11}.  
Recently, Hirunsit {\sl et. al}.\cite{12} studied the influence of Al, B, N and P doping on graphene electronic structures and change in quantum capacitance by using DFT calculations. Their report indicates that the $C_Q$ in the monolayer graphene changes substantially when doped with N and in presence of vacancy defect. Later, Song {\sl et al}\cite{13} studied the quantum capacitance in ad-atom functionalized reduced graphene oxide and found a significant enhancement in $C_Q$. Therefore it is not difficult to realize from the recent studies that the quantum capacitance in graphene based electrodes can be improve significantly by means of an adequate functionalization. Several study of $C_Q$ on functionalized graphene have been recently reported, however, a systematic investigation of quantum capacitance in functionalized graphene considering various type of ad-
atoms with a variable concentration and the basic theoretical understanding of their effect on the $C_Q$ is still lacking.
In this present study we have used density functional theory calculations to investigate the quantum capacitance of different functionalized graphene in a systematic way. The functionalization has been done using ad-atoms from different groups in periodic table. The role of vacancy defects on electronic structure and its effect on quantum capacitance in functionalized graphene(FG) has also been carefully investigated.

\section{Computational Method} 
To accomplish our theoretical investigation of $C_Q$, we first obtain the accurate electronic structure of the doped graphene using plane wave based density functional theory calculations implemented in  Vienna Ab-initio Simulation Package(VASP)\cite{14,15,16}. Projected augmented wave method\cite{17} was used to optimize the geometric structure of the functionalized graphene. The exchange correlation energy functionals were approximated using generalized gradient approximation with PBE parametrization\cite{18,19}. A very high kinetic energy cut-off (\textgreater 400eV) was used in all our calculations for the accurate results. In order to explore the effect of different functionalization on the quantum capacitance, calculations were done using 3$\times$3$\times$1 supercells of graphene unit cell, having 18 carbon atoms of graphene sheet (G18) with one functional group. The vacancy defected configurations were realized on a 5$\times$5$\times$1 supercell of graphene unit cell (50 C atoms of graphene, G50) with a variable concentration of vacancy in the range from 2 to 8 percent. A sufficiently large vacuum has been considered along the out of plane direction of graphene sheet (height$\textgreater$10\AA) to avoid the interaction with periodic images. We used a 6$\times$6$\times$1 $\Gamma$ point pack of k-point mesh to sample the Brillouin zone for geometry optimization with  10$^{-6}$H tolerance in total energy for convergence. A denser 24$\times$24$\times$1 k-point grid was used for the precise extraction of electron density of states D(E) and atom projected density of states(PDOS). 

The quantum capacitance of materials can be seen as the rate of change of excessive charges(ions) with respect to the change in applied potential\cite{20}. Therefore, it is directly related to the electronic energy configuration of the electrode materials and can defined as the derivative of the net charge on the substrate/electrode with respect to electrostatic potential. i.e.
\begin{equation}
C{_Q}$ = $\frac{dQ}{d\phi}
\end{equation}

 where Q is the excessive charge on the electrode and $\phi$ is the chemical potential.
 The total charge is proportional to the weighted sum of the electronic density of states upto the Fermi level $E_F$.  Due to an applied potential, the chemical potential will be shifted, the excessive charge on the electrode (Q) then can be expressed by an integral term associated to the electronic density of state D(E) and the Fermi$-$Dirac distribution function f(E) as
 
 \begin{equation}
 Q = e\int_{-\infty}^{+\infty} D(E)[f(E) - f(E - \phi)] dE
 \end{equation}
 
  Therefore, when the density of states (DOS) is known, the quantum capacitance $C_Q$ of a channel at a finite temperature T can be calculated as

\begin{equation}
C{_Q} =\frac{dQ}{d\phi} = \frac{e^2}{4kT}\int_{-\infty}^{+\infty} D(E) Sech^{2}\left(\frac{E - {e\phi}}{2kT}\right) dE
\label{qc}
\end{equation}

Here {\sl e} is the electrons charge,  $\phi$ is the chemical potential, {\sl D(E)} is the DOS and $k$ is the Boltzmann constant.
We therefore  have estimated the $C{_Q}$ for all the system directly form the density of states. 

\section{Results and discussion}

\begin{figure}[!ht]
\centering
  \includegraphics[width=0.5\linewidth]{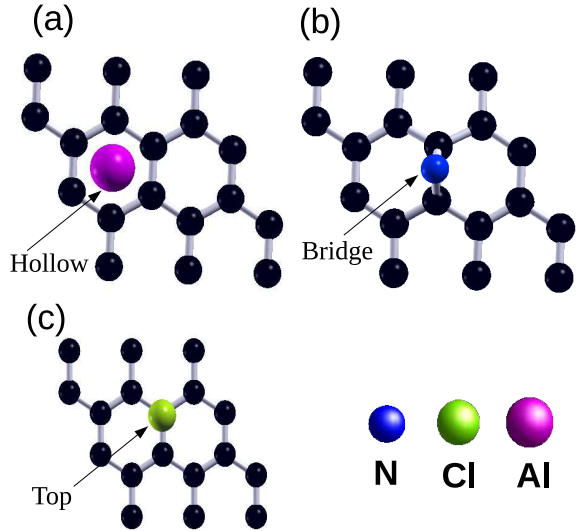}\quad
\caption{(colour online) Energy-optimized geometry of functionalized graphene with different ad-atoms. Black magenta, blue and green ball represents C, Al, N and Cl atoms respectively a) Preferred adsorption position for Al at hollow site b)Nitrogen at bridge site and (c) Chlorine at the top site. }
\label{FG-Adatom Optimized structure}
\end{figure}

It is evident from the expression of $C_Q$ in equation (\ref{qc}) that the quantum capacitance is directly proportional to the density of state present near the Fermi energy. Since $(E-e\phi)$ represents the energy with respect to Fermi level and $Sech^{2} (x)$ rapidly goes to zero for $|x| > 0$, the states which are energetically far from the $E_F$ are not contribute much on $C_Q$ . The density of state near the Fermi level for a given material can be tune by means of an efficient chemical modification of the system using external ad-atoms or creating defects. This is also an effective way to control the type and concentration of charge carriers in the system. The electronic Energy levels of the parent material may also be modified/shift in these process. The change in electronic structure depends on dopant type, concentration and doped position in the sublattice. In this study we have considered atoms from different groups in periodic table with an increasing order of electronegativity such as K$<$ Na$<$ Al$<$ P$<$ N$<$ Cl, to functionalize the graphene. The stable adsorption position  on graphene was estimated by placing ad-atoms in different possible adsorption sites and comparing the optimized total energies. The hollow position was found to be the most favourable position for ad-atoms like K, Na, Al, bridge position for  P, N and top position for Cl ad-atoms respectively. The optimized structure of functionalized graphene with different adsorption positions are shown in Fig. \ref{FG-Adatom Optimized structure}. The stability of the functionalized structure was investigated by estimating average adsorption energy $E_{ad}$ for ad-atoms using 
\begin{equation}
E_{ad} = \frac{1}{n}[E_{tot} - E_{gr} -nE_{at}]
\end{equation}
where $E_{tot}$ is the total energy of the functionalized graphene unit cell,   $E_{gr}$ is the total energy of pristine  graphene in the same unit cell, $E_{at}$ is the per atom energy of the ad-atom and $n$ represents the number of ad-atoms present in the unit cell. The adsorption energies for different ad-atoms are listed in the Table. \ref{table1}. Our calculation shows that the adsorption of N on pristine graphene surface is most favourable compare to other ad-atoms. However relatively large  adsorption energies for P, Cl and Al clearly indicate that these can be easily adsorb on the pristine graphene surface. The ad-atoms are adsorbs on graphene at a distances varying from 1.5 \AA \ to 3.3 \AA \ from the surface. Interestingly, the planar structure of graphene is not much disturbed with these adsorptions. However a very small change in the C-C bond lengths (order of 0.002nm) compare to C-C bond length of graphene were observed near the doped site.

\begin{table}
\caption{\label{table1}Adsorption energy per ad-atoms adsorbs on pristine graphene surface.}
\begin{indented}
\item[]\begin{tabular}{@{}llll}
\br
\textbf {ad-atom} & \textbf{ adsorption}& \textbf{ad-atom} & \textbf{ adsorption}\\
                  & \textbf{energy(in eV)}&               & \textbf{energy(in eV)}\\ 
\mr
Na & -0.525   & Cl & -1.142 \\
K  & -0.790   & P  & -1.234 \\
Al & -1.117   & N  & -2.372 \\ 
\br
\end{tabular}
\end{indented}
\end{table}

The electronic structure and subsequently the quantum capacitance of functionalized graphene were calculated in the optimized adsorbed geometry. We observed a slight shifts of band energies in graphene due to functionalization. 
Our calculation shows that quantum capacitance in functionalized graphene varies with type, concentration and distributions of ad-atoms. The calculated  $C_Q$ values in functionalized graphene with ad-atoms from different groups with a 5.5\%  doping concentration are shown in Table. \ref{table2}, 
Our calculated result clearly indicates a significant enhancement in quantum capacitance of graphene due to functionalization in comparison with the pristine graphene which is $\sim 1.3 \mu F/cm^2 $. A similar trend is reported in previous studies\cite{20}. Notice  that enhancement of $C_Q$  is proportional to the increase of electronegativity of ad-atoms.

\begin{table}
\caption{\label{table2}Calculated $C_Q$ value at 300 K  for various ad-atom functionalized graphene with doping concentration 5.5\%.}
\begin{indented}
\item[]\begin{tabular}{@{}lll}
\br
\textbf {Configuration} & \textbf{ Electronegativity }& \textbf{C$_Q$ } \\
                  & \textbf{of ad-atom} & \textbf{($\mu$F/cm$^2)$}            \\ 
\mr
 Pristine Graphene  & - & 1.2947 \\
  FG - K   &  0.82  &  26.7905     \\ 
  FG - Na  &  0.93  &  75.9058     \\ 
  FG - Al  &  1.61  &  57.4976      \\
  FG - Sn  &  1.96  &  249.4770    \\
  FG - P   &  2.19  &  346.1162     \\
  FG - N   &  3.04  &  256.4227    \\
  FG - Cl  &  3.16  &  553.6849    \\
\br
\end{tabular}
\end{indented}
\end{table}

The doped atoms in our study can be broadly categorized in two different groups.  Group 1-3 metals such as K, Na and Al ad-atoms are  behave as electron donors for graphene. They are donating electrons from  their outer shell to the graphene and shift the Fermi level of graphene into the conduction band. As a result the system shows a n-type behaviour. On the other hand group 15 and 17 elements such as N and Cl are electron accepting ad-atoms (p-type doping).

\begin{figure}[!ht]
\centering
 \includegraphics[width=0.7\linewidth]{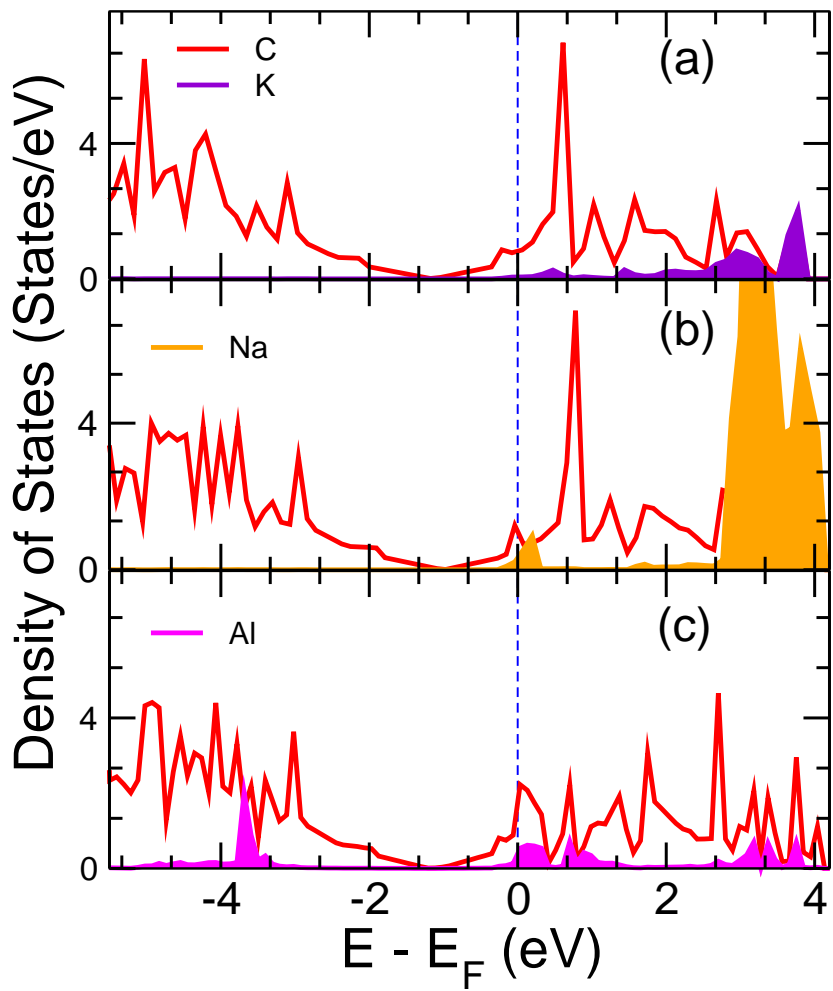}\quad
\caption{(colour online) Atom projected density of states for functionalized graphene with (a)K (b)Na and (c)Al atoms. The shaded curve represents dos from the doped atom. The vertical blue dashed line is the Fermi energy set at E=0.}
\label{n-type doping-PDOS}
\end{figure}

In case of n-type functionalization, the Dirac cone structure of the graphene bands are preserved as shown in  Fig. \ref{n-type doping-PDOS}. The Fermi level is located at the conduction cone, indicating a large amount of ad-atom induced free electron density between the Dirac point energy (E$_D$) and E$_F$, which can be controlled by tuning the n-type ad-atom concentration. 

\begin{figure}[!ht]
  \centering
 \includegraphics[width=0.7\linewidth]{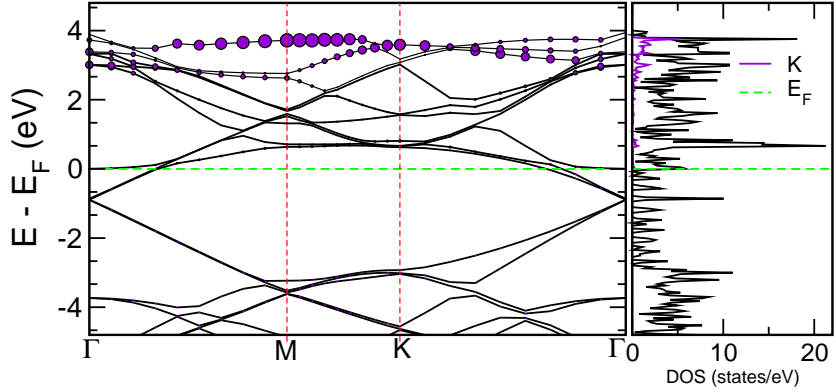}\quad
\caption{(colour online)Electronic band structure and DOS for graphene functionalized with K. Contribution from doped atoms are represented by the coloured curve in the DOS and coloured circle in the band structure. Horizontal green dashed line is the Fermi energy set at E=0.}
\label{BS-FG-K}
\end{figure}

 The density of states contributed from  alkali metal ad-atoms, are far from Fermi energy and shows a weak dispersion. K doped graphene band structure is shown in Fig. \ref{BS-FG-K}, where  coloured circles represent band contributed from the doped K atoms. Since the conduction in pristine graphene is mainly due to the de-localized $\pi$ cloud from p$_z$ orbitals of carbon, only p-states are expected to be present around the Fermi-energy. A very similar change in electronic structure were observed for Na and Al doped graphene shown in Fig. \ref{n-type doping-PDOS}(b)\&(c). The overall change in density of state near the Fermi energy very small. As a result, the change in quantum capacitance compared to pristine graphene is expected to be very small. The maximum value of C$_Q$  among all alkali metal doped graphene is found to be $\sim 76 \mu F/cm^2$ for Na doping.
 
\begin{figure}[!ht]
	\centering
 \includegraphics[width=0.7\linewidth]{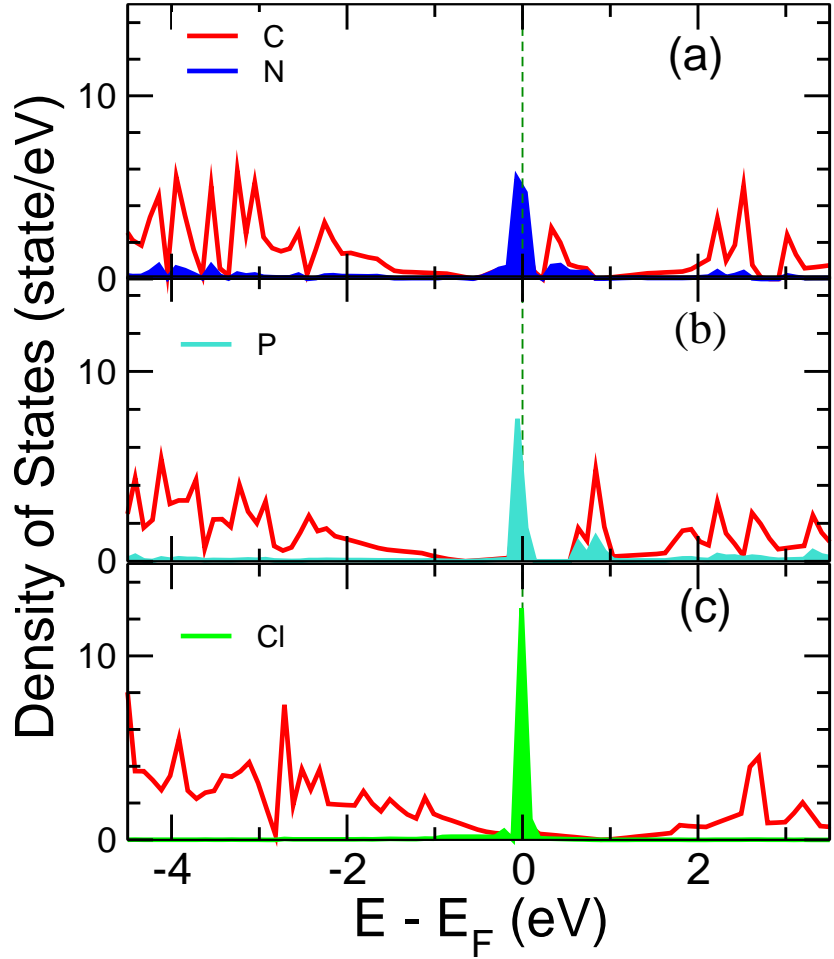}\quad
\caption{(colour online)Atom projected density of states for functionalized graphene with  (a)N (b)P and (c)Cl ad-atoms. The shaded curve represents dos from the doped atom. The vertical blue dashed line is the Fermi energy set at E=0. }
\label{p-type doping-PDOS}
\end{figure}

\begin{figure}[!ht]
	\centering
	\includegraphics[width=0.85\linewidth]{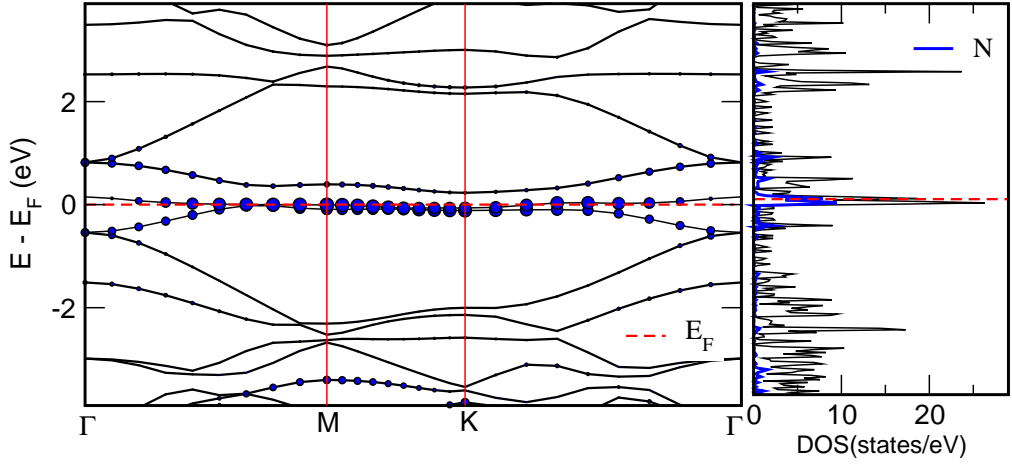}\quad
	\caption{(colour online)Electronic band structure and DOS for graphene functionalized with N. Contribution from doped atoms are represented by the coloured curve in the DOS and coloured circle in the band structure. Horizontal red dashed line is the Fermi energy set at E=0.}
	\label{BS-FG-N}
\end{figure}

The electronic behaviour of p-doped graphenes  are entirely different from the alkali metal doped graphene. 
In all of p-doped graphene system, Dirac cone structure got distorted and the Fermi level is located at the valence Dirac cone. The atom projected density of states for N, P and Cl doped graphene are shown in Fig. \ref{p-type doping-PDOS}. Free hole density appear between E$_D$ and E$_F$. The energy states near the Fermi energy are mainly contributed from the dopant atoms. We observed a maximum change is for Cl doped system. It shows strong peaks near $E_F$ which are $3p_z$ states from the doped chlorine ad-atom as shown in  Fig. \ref{p-type doping-PDOS}. N doped graphene band structure is shown in  Fig. \ref{BS-FG-N}, where  coloured circles represent band contributed from the doped N atoms.

 Accumulation of large density of state near the Fermi energy was found in all p-doped graphene. Since the quantum capacitance is directly proportional to the measure of electron density near Fermi level, there is a significant enhancement observed in case of p-type  doping system  shown in Table \ref{table2}. We found the maximum value of $C_Q = 553 \mu F/cm^2$ for Cl-functionalized graphene with a 5.5\% doping concentration. 

\begin{figure}[!ht]
  \centering
  \includegraphics[width=0.6\linewidth]{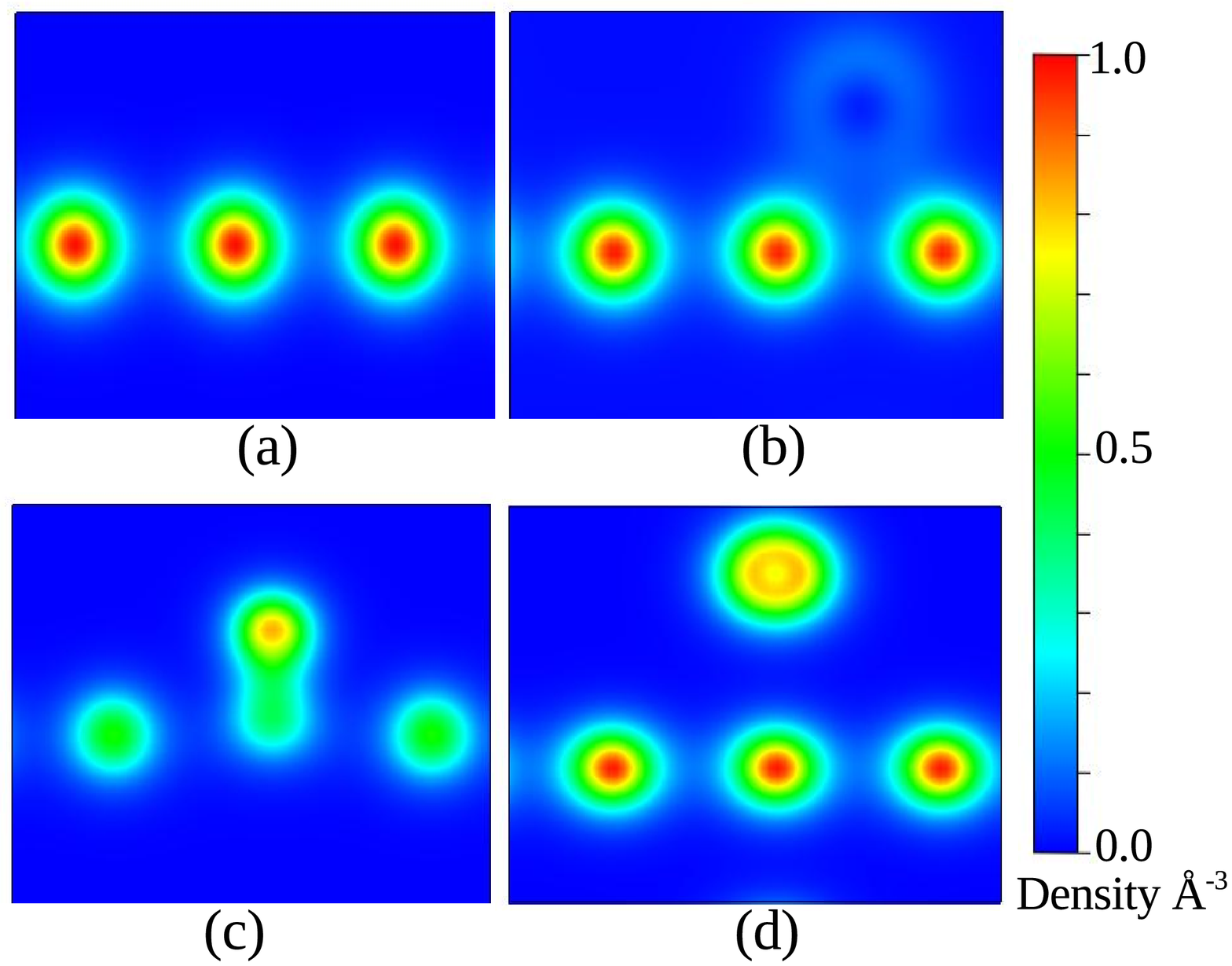}
\caption{(colour online)  Contour plots for electron density associated with (a)Pristine Graphene, Functionalized graphene with (b)Al, (c)N and (d)Cl. Relative electron density is indicated by the colour bar.}
\label{Isosurface-FG-Adatom}
\end{figure}

\begin{figure}[!ht]
	\centering
  \includegraphics[width=0.7\linewidth]{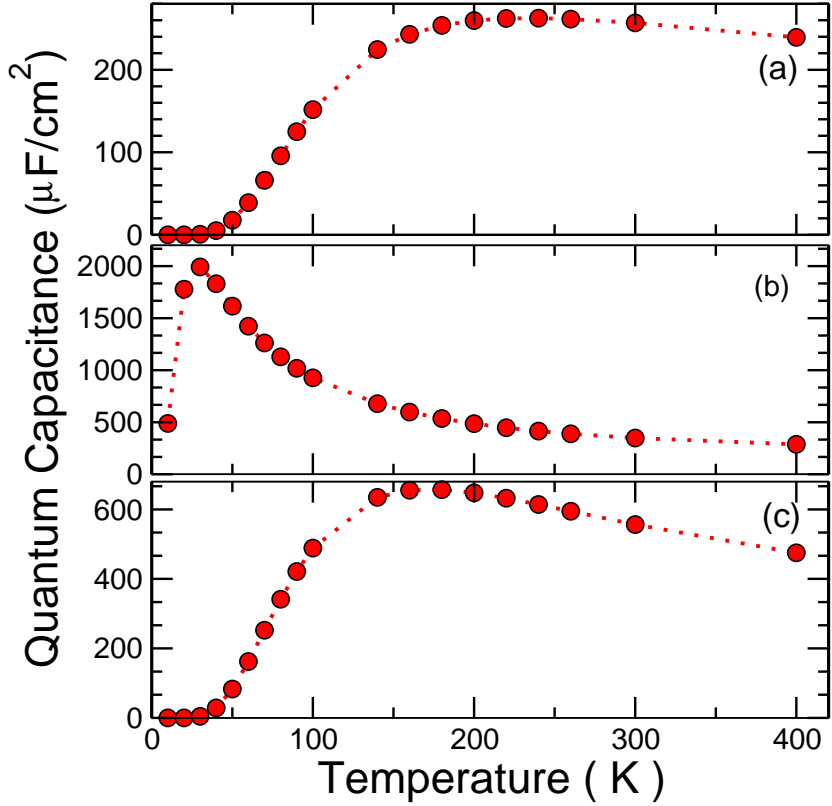}
\caption{(colour online) Variation of Quantum Capacitance with temperature in the range of 10K to 400K for (a)FG-N (b) FG-P and (c)FG-Cl.}
\label{QC-T}
\end{figure}

Next we investigated the origin of large enhancement in quantum capacitance in p-type doping on graphene. Charge redistribution in presence of ad-atom could be one of the main reason. Therefore, we performed Bader charge analysis on functionalized graphene and observed that there are significant charge transfer between graphene and the ad-atoms.
 In case of Cl and N doping, the charge transfer is remarkably high. We found 0.6e and 0.5e charge par atom transfer to Cl and N atoms respectively from the graphene sheet. In case of P functionalized graphene, 0.3e charge has been transferred from P to the graphene sheet. A small change in the sublattice structures have also been observed in all case. The introduction of the electron accepting(donating) ad-atoms, disrupts the homogeneity of the charge distribution due to the strong correlation effects. 
 Fig. \ref{Isosurface-FG-Adatom}  shows the charge redistribution upon functionalization of graphene with Al, N and Cl ad-atoms. A uniform  charge distribution can be seen in the case of pristine graphene monolayer (See Fig. \ref{Isosurface-FG-Adatom}(a)).
We notice that the charge inhomogeneity systematically increases with increasing atomic number of the ad-atom. The localization of charges near the dopant site increases systematically as we move from the Na to Cl. This can be ascribed to the fact that the strong on-site Coulombic interaction dominate in case of heavier p-block elements. We found maximum charge re-distribution on graphene caused by Cl. Further charge transfer leads to a change in the average DOS near Fermi level, which in-turn affect the quantum capacitance. Interestingly the density of states from  $3p_z$ orbitals of Cl, N and P are localized very near to Fermi level as shown in Fig. \ref{p-type doping-PDOS}, which provides the maximum contribution to the quantum capacitance. To understand the localization of states near the Fermi energy from the doped atom we investigate the temperature dependent behaviour of the $C_Q$.
We vary the temperature from 10K to 400K and calculated the $C_Q$ of the system. In case of pristine graphene monolayer the value of $C_Q$ remains nearly unchanged and it is close to $\sim 1.3 \mu$F/cm$^2$.
However, $C_Q$ changes dramatically with temperature when Cl, P and N atoms are added to the graphene.
Our calculation shows a sharp increase and then a gradual decrease of $C_Q$, when we increase the temperature. With a 5.5\% doping concentration the maximum value of $C_Q= 648 \mu F/cm^2$ appears at 200K in Cl doped graphene and  263 $\mu F/cm^2$ arises at 234K in N-doped graphene. A significantly large value of $C_Q=1992 \mu F/cm^2 $ was observed for P doped system at 30K with a same doping concentration. The variation of $C_Q$ with temperature for N, P and Cl doped graphene are shown in Fig. \ref{QC-T}. One possibility of such variation of $C_Q$ with temperature could be due to the Kondo behaviour of the doped graphene where ad-atoms may behave as the magnetic impurities. To confirm this, we perform spin polarized DFT calculations for these systems and observed that 0.371$\mu B$, 0.75$\mu B$, and 0.45 $\mu B$ magnetic moment lies on the doped Cl, N, P atoms respectively, which confirms the magnetic behaviour of the doped ad-atoms and localization of DOS near the Fermi energy.

\subsection{Dependence of ad-atom concentrations on $C_Q$:}

 Next we investigated the impact of doping concentration on quantum capacitance. We choose Cl, N and P and their combination as ad-atoms to dope graphene with a variable concentrations, as we observed that doping with these atoms with a similar concentration shows comparatively large $C_Q$ value at the room temperature. 
 
 We observed a steady increment of C$_Q$ for N and Cl doped system  upto a certain concentration and then it decrease. However, in case of P doped system the C$_Q$ decreases significantly with increase of doping concentration also we found that the P adsorption is not stable when the doping concentration is more than 8\%. The maximum value of C$_Q$ appears to be 284, 280 and 1142  $\mu F/cm^2 $ for P, N and Cl doped systems with doping concentration 2\%, 6.25\% and 12\% respectively at room temperature. The concentration dependence of $C_Q$ for different systems are summarized in the Table \ref{table3}. We also observed that the $C_Q$ is sensitive to the adsorption site of the graphene lattice and it reduces drastically when ad-atoms are very close to each other. Moreover the calculated value of $C_Q$ is very small, when more than one ad-atom adsorbs on a single ring unit of the graphene or in equivalent sites of the adjacent ring. Since the N is prefer to adsorb on the bridge position and Cl prefers to adsorb on top position of the graphene lattice, therefore maximum concentration with such restriction is possible only for 6.25\% and 12\% doping concentration for N and Cl adsorption respectively. When the doping concentration is more than the critical value, impurity centres are interacting each other that affects on the localized states near the Fermi energy, which inturn reduces the $C_Q$ value.

 \begin{table}
\caption{\label{table3}Concentration dependence in C$_Q$ for nitrogen and chlorine functionalized graphene at 300K.}
\begin{indented}
\item[]\begin{tabular}{@{}llll}
\br
\textbf {Concentration of ad-atom} & \textbf{FG-N:C$_Q$}& \textbf{FG-Cl:C$_Q$} & \textbf{FG-P:C$_Q$}\\
                  & \textbf{($\mu$F/cm$^2$ )}& \textbf{($\mu$F/cm$^2$ )}& \textbf{($\mu$F/cm$^2$ )} \\  
\mr
 2\%      &  254.344   &  315.7274  & 284.9665 \\  
  4\%      &  269.484   &  1066.193 & 279.3714 \\ 
  6.25\%   &  279.770   &  976.1142 & 281.5697   \\ 
  8\%      &  249.099   &  980.3039 & 108.6805 \\ 
  10\%     &  195.3198  &  279.7531 &    \\ 
  12\%     &  269.7866  &  1141.6553 &   \\ 
\br
\end{tabular}
\end{indented}
\end{table}
 
%
%
 
\begin{figure}[!ht]
 	\centering
 	\includegraphics[width=10cm]{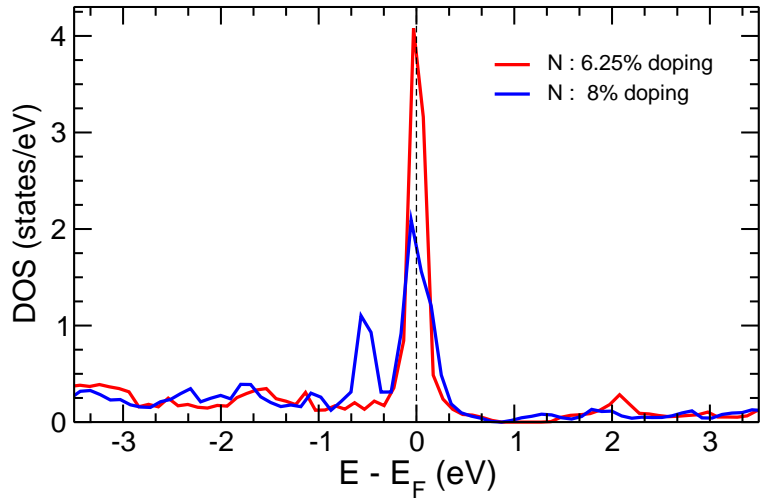}
 	\caption{(colour online) N atom projected DOS for nitrogenated graphene with two different N concentrations. Red line is for the 6.25\% doping concentration, where the $C_Q$ is maximum. Blue line represent the same for higher 8\% doping concentration. }
 	\label{Max-QC-Doping for FG-N}
 \end{figure}

  This can be understand by comparing the projected DOS of the impurity N atom in N-doped graphene with  critical doping concentration(red curve) and higher than the critical doping concentration (blue curve) as shown in Fig. \ref{Max-QC-Doping for FG-N}, where the impurity band width is extended for the concentration which is higher than the critical value. Dense decoration of Cl on a graphene surface leads to the desorption of Cl from the surface in the form of Cl$_2$ due to a stronger Cl$-$Cl interaction.
  
   A very similar situation appears in case of co-doping of Cl, N and P ad-atoms. In this case the interaction between ad-atoms are much more stronger and instabilizes the absorption of ad-atoms on the graphene surfaces. The $C_Q$ also reduces significantly as shown in the Table \ref{table4}.
   
   \begin{table}
\caption{\label{table4}Calculated $C_Q$ values for Co-doped graphene(G18) with two different ad-atoms with 5.5 \% doping concentration for each type.}
\begin{indented}
\item[]\begin{tabular}{@{}ll}
\br
\textbf {Co-doped ad-atoms} &
 		 \textbf{C$_Q$($\mu$F/cm$^2$)}   \\ 
\mr
 N,Cl   &  110.05 \\
 N,Sn   &  118.22 \\
 P,Sn   &  118.45 \\ 
 Sn,Cl  &  102.57  \\ 
\br
\end{tabular}
\end{indented}
\end{table}

\subsection{Impact of vacancy defects on $C_Q$}

Dislocation or defect on pristine graphene also causes charge localization, which inturn may improve the quantum capacitance. Therefore we have investigated the quantum capacitance of graphene in presence of vacancy defects. The vacancies were created by simply removing C atom from the graphene lattice.
 Our calculation shows that $C_Q$ increase with defect concentration up to 4\% and then it decrease. The enhancement of the quantum capacitance is due to the formation of localized states near Fermi energy, induced by defects, which is  very similar to the effect of ad-atom doping on graphene. It was also observed that the localized states are spin polarized induces $\sim 0.3 \mu B$ magnetic moment for each vacancy created. The value of $C_Q$ in different vacancy concentration are listed in the Table. \ref{table5}. We were unable to study the vacancy defected graphene with more than 8\% defect as the structure itself is unstable. Interestingly the position of the created vacancy site also has an impact on $C_Q$, very similar to the ad-atom's positions in doped graphene. We observed a significant reduction of $C_Q$ in 4\% defect configuration when the  distance between two defect sites is less than 5 \AA.

\begin{table}
\caption{\label{table5}Calculated $C_Q$ values for vacancy defected graphene with  different vacancy concentration(on G50) at 300K.}
\begin{indented}
\item[]\begin{tabular}{@{}ll}
\br
\textbf { Vacancy concentration(\%)} & \textbf{C$_Q$($\mu$F/cm$^2$)}   \\ 
\mr
  0 & 1.2947 \\
  2 &  99.0641  \\
  4 & 260.3285 \\
  6 & 244.5882  \\
  8 & 127.8805  \\
\br
\end{tabular}
\end{indented}
\end{table}

 In the final step, we have incorporated vacancy defects in  graphene functionalized with N, P, Cl ad-atoms. In presence of vacancy defect, we  observed a slight variation in C$_Q$ only for Cl doped system. In 2\% and 4\% chlorine concentration, when one vacancy was created in a 50 C atom unit cell of graphene, the $C_Q$ increases to 486 $\mu$F/cm$^2$ and 1091 $\mu$F/cm$^2$ respectively. Further increment in Cl doping concentration, the $C_Q$ reduces drastically. This is due to the strong interaction between the localized charge in the doped atom and in the defected site. We have not fund any considerable change in  C$_Q$ for  N and P doped graphene with vacancy defects.

\section{Conclusion}

In conclusion, our theoretical study  shows that the   quantum capacitance of graphene-based electrodes can be enhance significantly by introducing ad-atoms and vacancy defects in the graphene sheet.  
Our density functional theory based calculations indicate that the quantum capacitance in functionalized graphene is mainly depends on the sublattice positions and the concentration of ad-atoms. We observed that the enhancement is significant when the graphene was functionalized with N, Cl and P ad-atoms with a specific concentrations. These ad-atoms are behaving as magnetic impurities in the system, generates localized density of states near the Fermi energy. Atom projected density of states for these systems show that new DOS are appearing near the Fermi-energy which are mainly from   p-orbitals of ad-atoms, produces high charge(electron/hole) density near the Fermi level results to a very high quantum capacitance in the system. Our calculation also predicts that the Cl functionalization is most feasible for higher $C_Q$  when defect present in the graphene. Finally the temperature variation calculation shows that $C_Q$ remains large in case of Cl and N functionalized graphene in a wide range of temperature. Our study propose that the chemically modified graphene could be a promising electrode materials for super-capacitors, in which a very large quantum capacitance (($>$600 $\mu$F/c$m^2$)) can be achieve near the room temperature. 

\section{Acknowledgement}
KT would like to acknowledge NITK-high performance computing facility and also would like thank DST-SERB(project no. SB/FTP/PS-032/2014 ) for the financial support.

\Bibliography{99}

\item El-Kady, M. F., Shao, Y. and Kaner, R. B. 2016, {\it Nature Reviews Materials}, {\bf 1}(7), p. 16033. \\

\item Novoselov, K. S. {\it et al.} 2005, {\it Nature}, {\bf 438}(7065), pp. 197-200. \\

\item Zhang, L. L., Zhou, R. and Zhao, X. S. 2009, {\it Journal of Materials Chemistry}, {\bf 38}(29), pp. 2520-2531. \\

\item Ke, Q. and Wang, J. 2016, {\it Journal of Materiomics}. Elsevier Ltd,{\bf 2}(1), pp. 37-54.  \\

\item Zhan, C. {\it et al.} 2015, {\it Journal of Physical Chemistry C}, {\bf 119}(39), pp. 22297-22303.\\

\item Yu, G. L. {\it et al.} 2013, {\it Proceedings of the National Academy of Sciences of the United States of America}, {\bf 110}(9), pp. 3282-6. \\

\item Droscher, S. {\it et al.} 2010, {\bf 96}(15). \\

\item Mousavi-Khoshdel, S. M. and Targholi, E. 2015, {\it Carbon. Elsevier Ltd}, {\bf 89}, pp. 148-160. 
\\

\item Pak, A. J., Paek, E. and Hwang, G. S. 2014, {\it Carbon. Elsevier Ltd}, {\bf 68}(512), pp. 734-741.  \\

\item Zhan, C. {\it et al.} 2016, {\it Phys. Chem. Chem. Phys. Royal Society of Chemistry}, {\bf 18}(6), pp. 4668-4674.\\

\item Yang, G. M. {\it et al.} 2015, {\it Journal of Physical Chemistry C}, {\bf 119}(12), pp. 6464-6470.\\

\item Hirunsit, P., Liangruksa, M. and Khanchaitit, P. 2016, {\it Carbon. Elsevier Ltd}, {\bf 108}, pp. 7-20.\\

\item Song, C. {\it et al.} 2018, {\it ChemPhysChem}, {\bf 19}(13), pp. 1579-1583.  \\

\item Kresse, G. and Furthmüller, J. 1996, {\it Computational Materials Science}, {\bf 6}(1), pp. 15-50. \\

\item Kresse, G. and Furthmüller, J. 1996, {\it Phys. Rev. B. American Physical Society},{\bf 54}(16), pp. 11169–11186 \\.

\item Hohenberg, P. and Kohn, W. 1964, {\it Phys. Rev. American Physical Society}, {\bf 136}(3B), pp. B864--B871.

\item Blochl, P. E. 1994,  {\it  Phys. Rev. B. American Physical Society}, {\bf 50}(24), pp. 17953-17979.\\

\item Perdew, J. P. {\it et al.} 1992, {\it Phys. Rev. B. American Physical Society}, {\bf 46}(11), pp. 6671–6687. 

\item Perdew, J. P., Burke, K. and Ernzerhof, M. 1997, {\it Phys. Rev. Lett. American Physical Society}, {\bf 78}(7), p. 1396.\\

\item Mousavi-Khoshdel, M., Targholi, E. and Momeni, M. J. 2015, {\it Journal of Physical Chemistry C}, {\bf 119}(47), pp. 26290-26295. \\

\endbib

\end{document}